\documentstyle[elsart12,osa,epsf,eqsecnum]{revtex}
\begin{document}


\voffset1.5cm


\title{Jet Quenching versus Jet Enhancement: a quantitative\\ study
of the BDMPS-Z gluon radiation spectrum}
\author{Urs Achim Wiedemann}

\address{Theory Division, CERN, CH-1211 Geneva 23, Switzerland}

\date{\today}
\maketitle

\begin{abstract}
We study the gluon radiation spectrum off a hard in-medium produced
quark in the multiple soft rescattering formalism of
Baier-Dokshitzer-Mueller-Peign\'e-Schiff, and of Zakharov (BDMPS-Z).
Its dependence on the quark and gluon energy,
on the gluon transverse momentum, on the in-medium pathlength and
on the rescattering properties of the nuclear medium is analyzed
quantitatively. The two components of gluon radiation, the hard
vacuum radiation associated to the quark production vertex, and the
medium-induced rescattering contribution interfer destructively. 
For small spatial extensions of the medium, this
destructive interference overcompensates the hard vacuum radiation,
and the total medium-induced radiative energy loss decreases
as $\Delta E \propto -\, L^3$. Medium-induced gluon production
dominates only above a finite critical length $L > L_{\bf crit}$
which varies between 3 and more than 6 fm depending on the 
rescattering properties of the medium. Deviations from the
BDMPS-$L^2$-behaviour persist above $L_{\bf crit}$. The medium
dependence of the angular gluon distribution is dominated by
transverse Brownian ${\bf k}_\perp$-broadening. This results in
a depletion of the low transverse momentum part of both the hard and
the medium-induced contribution. As a consequence, the medium-induced
energy loss outside a finite angular cone size $\Theta$ can be more 
than a factor two larger than the total medium-induced radiative 
energy loss. We discuss implications of these results for the jet 
quenching signal in relativistic heavy ion collisions at RHIC and
LHC.
\end{abstract}


\section{Introduction}
\label{sec1}

It is a longstanding goal to establish observables
which characterize in a model-independent way the non-perturbative 
properties of a nuclear medium. The medium-dependence of perturbatively 
calculable ``hard probes'' is studied intensively to this end.
A rigorous example available in e-A and p-A collisions is the 
generalized factorization theorem~\cite{LQS94} of Luo, Qiu and Sterman 
(LQS) which relates the medium-dependence of a class of high-$k_t$ 
observables to a very small set of higher twist parton distribution 
functions of the cold target nucleus. For ultrarelativistic 
A-A collisions, we are not aware of a corresponding factorization 
theorem for observables which could be measured with existing or
designed experiments at RHIC or LHC. This makes it interesting
to search outside the framework of higher twist parton distributions 
for other, rather process-independent medium characterizations which
allow to distinguish experimentally between the non-perturbative 
properties of hot and cold nuclear matter. 

A good candidate for a one-parameter characterization of non-perturbative 
medium properties is the rescattering parameter $n_0\, C$ obtained in the
QCD dipole picture~\cite{Z96,BDMPS97,WG99,W00qcd}. It describes the 
average squared transverse momentum transfered from the medium to an 
ultrarelativistic hard parton within one unit of pathlength, $n_0\, C = 
\langle {\bf q}_\perp^2\rangle / L$. This parameter increases with 
increasing nuclear density $n_0$ and (as a consequence of the equipartition 
theorem, $\langle {\bf q}_\perp^2\rangle \propto T$) with increasing 
temperature. However, in order to characterize on the basis of $n_0\, C$ 
differences between hot and cold nuclear matter, one has to understand 
quantitatively the sensitivity of different observables to $n_0\, C$. 
To this end, we study here the dependence of the medium-induced one gluon 
radiation spectrum off a hard quark on the rescattering parameter $n_0\, C$. 

Medium-induced radiative energy loss and the associated 
phenomenon of ``jet quenching'' was proposed a decade ago by
Gyulassy and Wang~\cite{WG91,GW94} as a sensitive probe of the 
non-perturbative properties of excited nuclear matter.  
Baier, Dokshitzer, Mueller, Peign\'e and 
Schiff (BDMPS) established~\cite{BDMPS97}
that its dominant contribution comes from the rescattering of the 
emitted gluon and results in a quadratic $L^2$-increase of 
$\Delta E/E$. An equivalent~\cite{BDMS-Zak,W00qcd} path-integral 
formulation~\cite{Z96,Z98} 
was derived independently by Zakharov (Z) - we refer to the formalism 
as BDMPS-Z. Recently, the present author~\cite{W00qcd} derived 
the ${\bf k}_\perp$-differential extension (\ref{2.1}) of this 
path-integral formalism which determines the transverse momentum 
distribution of the emitted gluon. Equation (\ref{2.1}), the 
starting point of the present work, differs from other 
formulations~\cite{BDMS99,Z99} by a regularization 
prescription~\cite{WG99,W00qcd} which makes
it applicable to finite size media. In particular, (\ref{2.1}) shows 
in an opacity expansion~\cite{W00qcd} the correct interpolation 
between totally coherent and totally incoherent radiation spectrum 
for spatially finite media. As we recall in section~\ref{sec3}, 
the medium-dependence of (\ref{2.1}) is determined by the 
rescattering parameter $n_0\, C$ only.

Quantitatively, little is known about the one gluon radiation
spectrum in the soft multiple scattering approximation. For the 
total, ${\bf k}_\perp$-integrated energy loss $\Delta E$, numerical 
results by Zakharov~\cite{Zmoriond98} indicate weak deviations
from the BDMPS-$L^2$-law. However, a systematic study of its
dependence on the rescattering parameter $n_0\, C$ and the energy
$E$ of the incident quark is missing. Moreover, all existing
results on $\Delta E$ are obtained by integrating the transverse gluon 
momentum up to infinity. In particular, it remains unclear to what
extent deviations from the BDMPS-$L^2$-law become more significant
if the finite kinematical boundary of the ${\bf k}_\perp$-integration
and the proper regularization of the gluon radiation spectrum are
taken into account. Also, there are no results which establish to what
extent the main contribution to $\Delta E$ stems from the region
of small gluon energy $\frac{\omega}{E} = x \ll 1$ and small
transverse gluon momentum $|{\bf k}_\perp| \ll \omega$ to which
the multiple soft rescattering formalism applies. This
motivates a systematic study of the $x$- and ${\bf k}_\perp$-differential
gluon radiation spectrum. The only related quantitative study due to 
Baier, Dokshitzer, Mueller and Schiff~\cite{BDMS99} determines from the
${\bf k}_\perp$-differential radiation spectrum the phenomenologically
important fraction $R(\Theta)$ of the energy $\Delta E$ radiated
outside a cone of opening angle $\Theta$. However, also in this 
study, the $|{\bf k}_\perp|$-integration extends to infinity and 
the questions listed above remain to be answered. This is done 
in the present work which is organized as follows: 

In section~\ref{sec2}, we recall the main properties of the one-gluon
radiation spectrum derived in the soft multiple scattering approximation.
Section~\ref{sec3} collects analytical expressions obtained in the dipole 
approximation. Numerical
results for the gluon radiation spectrum are presented 
in section~\ref{sec4} and their implications are discussed 
in the Conclusions.

\section{The medium-induced radiation spectrum}
\label{sec2}

We are studying the medium-induced, ${\bf k}_\perp$-differential one 
gluon radiation spectrum off a hard quark. In the multiple soft 
rescattering formalism, it reads~\cite{W00qcd}
\begin{eqnarray}
  &&{d^3\sigma^{(*)}\over d(\ln x)\, d{\bf k}_\perp}
  = {\alpha_s\over (2\pi)^2}\, {1\over \omega^2}\,
    N_C\, C_F\, 
    2{\rm Re} \int_{z_-}^{z_+} dy_l  
  \int_{y_l}^{z_+} d\bar{y}_l\, 
  e^{-\epsilon |y_l|\, -\epsilon |\bar{y}_l|}
  \nonumber \\
  && \qquad \times 
  \int d{\bf u}\,   e^{-i{\bf k}_\perp\cdot{\bf u}}   \, 
  e^{ -\frac{1}{2} \int_{\bar{y}_l}^{z_+} d\xi\, n(\xi)\, 
    \sigma({\bf u}) }\,
  {\partial \over \partial {\bf y}}\cdot 
  {\partial \over \partial {\bf u}}\, 
  {\cal K}({\bf y}=0,y_l; {\bf u},\bar{y}_l|\omega) \, .
    \label{2.1}
\end{eqnarray}
Here, ${\cal K}$ denotes the two-dimensional path-integral
\begin{equation}
 {\cal K}({\bf r}(y_l),y_l;{\bf r}(\bar{y}_l),\bar{y}_l|\omega)
 \int {\cal D}{\bf r}
   \exp\left[ \int_{y_l}^{\bar{y}_l} d\xi
        \left(i\frac{\omega}{2} \dot{\bf r}^2
          - \frac{1}{2}  n(\xi) \sigma\left({\bf r}\right) \right)
                      \right]\, .
  \label{2.2}
\end{equation}
The medium-dependence of (\ref{2.1}) stems from the
factor $n(\xi)\, \sigma({\bf r})$, where $n(\xi)$ determines
the density of scattering centers in the medium. The dipole
cross section
\begin{equation}
  \sigma({\bf r}) = 2\, C_A\, \int 
  \frac{d\bbox{q}_\perp}{(2\pi)^2}\, |a_0(\bbox{q}_\perp)|^2\, 
  \left( 1 - e^{-i\bbox{q}_\perp\cdot {\bf r}}\right)
  \label{2.3}
\end{equation}
contains configuration space information on the strength of a 
single elastic differential scattering cross section $\propto 
|a_0(\bbox{q}_\perp)|^2$.

The spectrum (\ref{2.1}) describes:
i) Gluon radiation off a hard {\it nascent} quark ($* = nas$)
produced during the collision inside the medium at $z_- = 0$. 
This is the medium-dependence of the QCD analogue of the radiation 
spectrum associated to $\beta$-decay.
ii) Gluon radiation off an incoming quark ($z_- = -\infty$, $*=in$)
contained in the wavefunction of the projectile. This is the
medium-dependence of the Gunion-Bertsch radiation spectrum~\cite{GB82}.

In the configuration-space formulation (\ref{2.1}), the quark 
radiates a gluon at longitudinal position $y_L$ ($\bar{y}_L$) in 
the amplitude (complex conjugate amplitude). The 
$\epsilon$-regularization of this cross section does not commute 
with the $z_\pm \to \pm \infty$ limit which removes the cut-off 
of the longitudinal integration. To do this analytically, we
consider a medium of finite longitudinal extension positioned 
along the longitudinal axis between 0 and $L$,~\cite{WG99}. 
We split the longitudinal integrals into six parts
\begin{equation}
  \int\limits_{z_-}^{z_+} \int\limits_{y_l}^{z_+}
  = \int\limits_{z_-}^{0} \int\limits_{y_L}^{0}
  + \int\limits_{z_-}^{0} \int\limits_{0}^{L}
  + \int\limits_{z_-}^{0} \int\limits_{L}^{z_+}
  + \int\limits_{0}^{L} \int\limits_{y_L}^{L}
  + \int\limits_{0}^{L} \int\limits_{L}^{z_+}
  + \int\limits_{L}^{z_+} \int\limits_{y_L}^{z_+}\, .
  \label{2.4}
\end{equation}
The six corresponding contributions to the cross section are
labelled in an obvious way:
\begin{eqnarray}
 &&\frac{d^3\sigma^{(*)}}{d({\rm ln}x)\,d{\bf k}_\perp} =
 \frac{\alpha_s}{\pi^2}\, N_c\, C_F\, 
 \sum_{j=1}^6 I_j\, .
\label{2.5}
\end{eqnarray}
Explicit expressions for the $I_j$'s, in which the limits 
$z_\pm \to \pm \infty$, $\epsilon \to 0$ are taken, are listed
in Appendix~\ref{appa}, Eqs. (\ref{a.2})-(\ref{a.7}). They
are the starting point of the following discussion.
We recall several technical points:

1. \underline{Region of validity of (\ref{2.1})}\\
Eq. (\ref{2.1}) is obtained by assuming small angle multiple 
scattering for both the projectile quark and the radiated 
gluon~\cite{BDMPS97}. Its region of validity is limited to
$x \ll 1$ and $|{\bf k}_\perp| \ll \omega$. For the calculation
of the total radiative energy loss, integration over 
$x$ and ${\bf k}_\perp$ is required. As discussed below, the 
$x$-integration is not problematic since the integrand has its
main support at small $x$. The ${\bf k}_\perp$-integration, however,
has significant support in the infrared and the
ultraviolet regime.

2. \underline{Infrared behaviour of (\ref{2.1})}\\
For vanishing medium thickness, the gluon radiation (\ref{2.1}) 
off a nascent quark is proportional to $\frac{1}{{\bf k_\perp}^2}$,
and the ${\bf k}_\perp$-integration is logarithmically divergent
in the infrared. To subleading order $O(x)$, an additional term
$\exp\left[{-i \bar{q} (y_L-\bar{y}_L)}\right]$ appears in the
integrand of (\ref{2.1}), with $\bar{q} \propto m_q^2$ the quark 
mass~\cite{Z96,KST98,WG99}. This term provides a natural infrared 
cut-off. In the
following analysis, however, we shall not modify (\ref{2.1})
by terms subleading in $O(x)$. We shall rather subtract from 
(\ref{2.1}) the medium-independent hard radiation associated 
with the quark production in vacuum. The medium-dependent 
remainder turns out to be infrared finite. This allows us to give 
a numerical analysis of the medium-induced radiative energy loss
without depending on an infrared cutoff parameter.

3. \underline{Ultraviolet behaviour of (\ref{2.1})}\\
The ${\bf k}_\perp$-integration over (\ref{2.1}) is ultraviolet
divergent even if the medium-independent part is subtracted off.
This is no fundamental problem, since the transverse momentum
is kinematically bound by the total energy of the gluon,
$|{\bf k}_\perp| \in [0,\omega]$. It indicates, however, that
the total radiative energy loss calculated from (\ref{2.1}) is 
sensitive to the high transverse momentum modes $|{\bf k}_\perp|
= O(\omega)$ which lie outside the region of validity of its
derivation. This is discussed in section~\ref{sec4e}.

\section{Dipole approximation}
\label{sec3}

Two approximation schemes are known for the spectrum (\ref{2.1}):
the opacity expansion and the dipole approximation. In the first,
one expands the spectrum in powers of the density of scattering
centers~\cite{WG99,W00qcd,GLV99,GLV00}. This also gives access
to the radiation off a small finite number of scattering centers.
In the dipole approximation, one exploits that the integrand of
(\ref{2.1}) has its main support at small transverse distances
$r = |{\bf r}|$. This allows to write the dipole cross section
(\ref{2.3}) to logarithmic accuracy as
\begin{eqnarray}
  \sigma({\bf r}) = C\, r^2\, .
  \label{3.1}
\end{eqnarray}
This approximation is used in all discussions of 
BDMPS~\cite{BDMPS97,BDMS-Zak,BDMS99} and of 
Zakharov~\cite{Z96,Z98,Zmoriond98,Z99}. It allows to explore
the multiple soft rescattering regime in which Brownian motion
dominates~\cite{WG99}. Due to (\ref{3.1}), the path-integral 
in (\ref{2.1}) simplifies to that of a harmonic oscillator
\begin{eqnarray}
  {\cal K}_{\rm osz}\bigl({\bf y},y_l;{\bf u},\bar{y}_l|\mu\bigr) 
     &=& {A\over \pi\, i} \exp\left[iAB({\bf y}^2 + {\bf u}^2)
             -2\,i\,A\,{\bf y}\cdot{\bf u} \right]\, ,
  \label{3.2} \\
  A &=&  {\mu\Omega\over 2\, \sin(\Omega\, \Delta y)}\, ,\qquad
  B = \cos(\Omega\, \Delta y)\, ,
  \label{3.3}
\end{eqnarray}  
with complex oscillator frequency
\begin{equation}
  \Omega = \frac{1-i}{\sqrt{2}}\, \sqrt{n_0\, C\over \omega}\, .
  \label{3.4}
\end{equation}
Here, the dipole strength $C$ appears multiplied by the homogeneous
density $n_0$ of scattering centers, $n_0\, C$. This is the only
model parameter in the following discussion.

In Appendix~\ref{appa}, Eqs. (\ref{a.8})-(\ref{a.13}), we write
the six contributions $I_j$ to the gluon radiation spectrum (\ref{2.5}) 
in the dipole approximation. For numerical evaluation, 
the integrands are expressed in dimensionless variables by
rescaling the transverse gluon energy and the longitudinal 
distance with the modulus of the oscillator frequency 
$\vert\Omega\vert$,
\begin{eqnarray}
 \tilde{\bf k}_\perp^2 &=& \frac{{\bf k}_\perp^2}{2\omega} 
 \frac{1}{\sqrt{2} |\Omega|}\, ,\qquad
 \tilde{z} = \sqrt{2} |\Omega|\, z\, ,\qquad
 \tilde{L} = \sqrt{2} |\Omega|\, L\, .
 \label{3.5}
\end{eqnarray}
For radiation off a nascent quark, the contributions $I_1$,
$I_2$, $I_3$ to (\ref{2.5}) vanish. The remaining three terms
can be represented by the diagrams
\begin{equation}
\epsfxsize=7.0cm 
\centerline{\epsfbox{ii456pap.epsi}}
\label{3.6}
\end{equation}
They denote the {\it direct production term} $I_4$ for gluon emission
inside the target in both amplitude and complex conjugate 
amplitude, the {\it destructive interference term} $I_5$ which is the
characteristic interference term between gluon emission in and 
beyond the target, and the {\it hard vacuum radiation term} $I_6$. 
The analytical expressions read (see
Appendix~\ref{appa} for intermediate steps)
\begin{eqnarray}
  I_4 &=& \frac{1}{2\, \omega\, \sqrt{2}\, |\Omega|}\,   
  2{\rm Re} \int_0^{\tilde{L}} d\tilde{z}_1
  \int_0^{\tilde{z}_1} d\tilde{z}_2\, 
  \exp\left[-\frac{2\tilde{\bf k}_\perp^2\, 
                   \sinh\left( \frac{1+i}{2}(\tilde{z}_1-\tilde{z}_2)\right)}
                 {N}\right]
  \nonumber \\
  && \qquad \qquad \times
  \left[ \frac{-i\, \tilde{z}_2}{N^2} -
         \frac{(2+2i)\, \tilde{\bf k}_\perp^2\,
               \cosh\left( \frac{1+i}{2}(\tilde{z}_1-\tilde{z}_2)\right)}
             {N^3}
               \right]\, ,
  \label{3.7}\\
  N &\equiv& \tilde{z}_2\, 
        \sinh\left( \frac{1+i}{2}(\tilde{z}_1-\tilde{z}_2)\right)
         + (1-i)\, 
        \cosh\left( \frac{1+i}{2}(\tilde{z}_1-\tilde{z}_2)\right)\, ,
  \label{3.8}\\
  I_5 &=& \frac{1}{2\, \omega\, \sqrt{2}\, |\Omega|}\,   
  2{\rm Re} \int_0^{\tilde{L}} d\tilde{z}
  \frac{-i}{\cosh^2\left( \frac{1+i}{2}\tilde{z}\right)}
  \nonumber \\
  && \qquad \qquad \qquad \times 
  \exp\left[ -(1+i)\, \tilde{\bf k}_\perp^2\, 
       \tanh\left( \frac{1+i}{2}\tilde{z}\right)\right]\, , 
  \label{3.9}\\
  I_6 &=& \frac{1}{{\bf k}_\perp^2}\, .
  \label{3.10}
\end{eqnarray}
The contribution 
$I_6$ is exactly (i.e. without taking recourse to the dipole
approximation) the medium-independent contribution associated to the
hard radiation off a nascent quark jet propagating in the 
vacuum \cite{W00qcd}. This follows from the leading $O(x)$
approximation in which only the rescattering of the radiated gluon 
contributes. Subtracting the medium-independent term $I_6$, the 
medium-induced modification to the hard radiation spectrum is given 
by $I_4$ and $I_5$,
\begin{eqnarray}
 &&\frac{d^3\sigma^{(nas)}_{med}}{d({\rm ln}x)\,d{\bf k}_\perp}
   =
 \frac{\alpha_s}{\pi^2}\, N_c\, C_F\, 
 \left( I_4 + I_5\right)\, .
 \label{3.11}
\end{eqnarray}
The radiation spectrum (\ref{2.5}) is positive, of course, since
one cannot radiate less than no gluon. This positivity is
guaranteed by the sum $I_4 + I_5 + I_6 > 0$ which according to
(\ref{3.6}) forms a complete square. For the difference (\ref{3.11})
between the radiation spectrum (\ref{2.5}) and the corresponding
vacuum contribution, positivity is not warranted: if less gluons
are emitted from the nascent off-shell quark in the medium than 
in the vacuum, then the medium-induced deviation turns negative.
The interplay between negative ({\it jet enhancement}) 
and positive ({\it jet quenching}) values of (\ref{3.11}) 
will be studied in what follows. 

\section{Numerical Results and Analytical Approximations}
\label{sec4}

The ${\bf k}_\perp$-integration of
(\ref{3.7})-(\ref{3.9}) is infrared finite. Using
polar coordinates
\begin{equation}
  \bar{I}_j = \int_0^{2\pi} d\varphi_k 
              \int_0^{\chi\, \omega} k_\perp\, dk_\perp
              I_j\, ,
              \label{4.1}
\end{equation}
one finds
\begin{eqnarray}
  \bar{I}_4 &=& 4\pi {\rm Re}  \int_0^{\tilde{L}} d\tilde{z}_1
                \int_0^{\tilde{z}_1} d\tilde{z}_2
                \left( \frac{i}{2} \frac{1}{1-\cosh[(1+i)\Delta \tilde{z}]}
                \right.
                \nonumber \\
            &&  \qquad \qquad + \left. 
                \frac{i}{4} \, 
                e^{- 2M_c^2 \sinh[(1+i)\Delta \tilde{z}/2] / 2}
                \, \frac{F}{N^2}
                \right)\, ,
                \label{4.2} \\
           F  &\equiv& 
                \tilde{z}_2^2 - 2i\coth[(1+i)\Delta \tilde{z}/2]
                \left[ (1+i)(M_c^2+\tilde z_2) 
                  \right.
                \nonumber \\
                && \qquad \qquad
                + \left. \coth[(1+i)\Delta \tilde{z}/2]\right]\, ,
                \label{4.3}\\
  \bar{I}_5 &=& 4\pi {\rm Re}  \int_0^{\tilde{L}} d\tilde{z}
                \left( \frac{-1}{2} \frac{1+i}{\sinh[(1+i)\tilde{z}]}
                  \right.
                  \nonumber \\
                && \qquad \qquad + \left.
                \frac{(1+i)\, e^{-(1+i)M_c^2 \tanh[(1+i)\tilde{z}/2]}}
              {4\, \sinh[(1+i)\tilde{z}/2] \cosh[(1+i)\tilde{z}/2]}\right)\, .
            \label{4.4}
\end{eqnarray}
Here, $\chi \in [0,1]$ denotes the fraction of the maximal transverse
momentum up to which we integrate, and $M_c$ denotes the corresponding
upper integration limit $\chi\, \omega$ in dimensionless rescaled variables
\begin{equation}
  M_c = \frac{\chi\, \omega}{\sqrt{ 2\omega \sqrt{2} |\Omega|}}\, .
  \label{4.5}
\end{equation}
These expressions determine the $x$-differential
radiation spectrum 
\begin{eqnarray}
  \frac{d\sigma^{(nas)}_{med}}{d({\rm ln}x)}(n_0\, C,L,E,\chi)
  &=& \int_{|{\bf k}_\perp|\leq \chi\omega}  d{\bf k}_\perp
        \frac{d^3\sigma^{(nas)}_{med}}{d({\rm ln}x)\,d{\bf
        k}_\perp} 
        \nonumber \\
  &=& \frac{\alpha_s}{\pi^2}\, N_c\, C_F\, 
 \left( \bar{I}_4 + \bar{I}_5\right)\, ,
 \label{4.6}
\end{eqnarray}
which is a function of the rescattering parameter $n_0\, C$, the 
pathlength $L$ of the quark in the medium, the energy of the quark
$E$ and the kinematical cut-off $\chi$. The $\chi$-dependence of 
(\ref{4.6}) allows to extract ${\bf k}_\perp$-differential information 
from the ${\bf k}_\perp$-integrated expression (\ref{4.6}), see
section~\ref{sec4e} below. The $x$-integrated 
total radiative energy loss $\Delta E$ is given by
 \begin{eqnarray}
        \frac{\Delta E}{E}(n_0\, C,L,E,\chi) = \int_0^1 dx\, x\, 
        \frac{d\sigma^{(nas)}_{med}}{dx}
        \label{4.7}\, .
 \end{eqnarray}
%
\subsection{Analytical Expansion for small $L$}
\label{sec4a}

\subsubsection{Radiation off projectile quark}
\label{sec4a1}

For the gluon radiation off an incoming quark, a Taylor expansion of 
the radiation spectrum (\ref{2.1}) around $\tilde{L}=0$ does not
exist~\cite{WG99}. One finds 
\begin{eqnarray}
 &&\frac{d^3\sigma^{(in)}}{d({\rm ln}x)\,d{\bf k}_\perp} =
 \frac{\alpha_s}{\pi^2}\, N_c\, C_F\, 
 \left(\int \frac{d^2{\bf q}_\perp}{(2\pi)^2}\, 
          \left(2\pi\over n_0\, C\, L\right)\, 
          e^{-\frac{{\bf q}_\perp^2}{2\, n_0\, C\, L}} \right.
          \nonumber \\
  && \times \left.
      \left[ \frac{1}{({\bf q}_\perp-{\bf k}_\perp)^2}
           + 2\frac{({\bf q}_\perp-{\bf k}_\perp)\cdot {\bf k}_\perp}
                   {({\bf q}_\perp-{\bf k}_\perp)^2\, {\bf k}_\perp^2}
           + \frac{1}{{\bf k}_\perp^2} \right]
                \right) 
           \left( 1 + O(L^2)\right)  + O(L^2)\, ,
\label{4.8}
\end{eqnarray}
which is non-analytic in $L$. 
The three contributions in the bracket [] of (\ref{4.8}) 
combine to 
\begin{equation}
\propto \int \frac{d^2{\bf q}_\perp}{(2\pi)^2}\, 
          \left(2\pi\over n_0\, C\, L\right)\, 
          e^{-\frac{{\bf q}_\perp^2}{2\, n_0\, C\, L}}
          \frac{{\bf q}_\perp^2}
                   {({\bf q}_\perp-{\bf k}_\perp)^2\, {\bf k}_\perp^2}\, .
          \label{4.9}
\end{equation}
This is the Gunion-Bertsch gluon radiation spectrum~\cite{GB82} 
for a target momentum transfer ${\bf q}_\perp$ which is acquired 
by the transverse Brownian motion of the incoming projectile,
\begin{equation}
  \langle {\bf q}_\perp^2 \rangle = 2\, n_0\, C\, L\, .
  \label{4.10}
\end{equation}
The dipole approximation does not break down for $L \to 0$
where the multiple soft rescattering formalism is questionable. 
Taking $L\to 0$ for $n_0\, C\, L = {\rm fixed}$, (\ref{4.9}) 
matches smoothly onto the single Gunion-Bertsch rescattering 
result~\cite{W00qcd}.

\subsubsection{Quark production in the medium}
\label{sec4a2}

The medium-dependence (\ref{4.6}) of the gluon radiation spectrum 
off a nascent quark can be expanded in a Taylor series for small
in-medium pathlength.
We find  
\begin{eqnarray}
  \bar{I}_4 &=& \frac{1}{8} \tilde{L}^2\, M_c^4 +
                \frac{1}{24} \tilde{L}^3\, M_c^2 + O(\tilde{L}^4)\, ,
                \label{4.11}\\
  \bar{I}_5 &=& \frac{-1}{8} \tilde{L}^2\, M_c^4 +
                 \frac{-1}{12} \tilde{L}^3\, M_c^2 + O(\tilde{L}^4)\, .
                \label{4.12}
\end{eqnarray}
Here, the destructive interference term $\bar{I}_5$ overcompensates
for thin media the direct production term $\bar{I}_4$. The resulting 
medium-induced deviation from the jet energy loss in the vacuum is 
negative,
\begin{equation}
  \bar{I}_4 + \bar{I}_5 = \frac{-1}{24} \tilde{L}^3\, M_c^2 
                          + O(\tilde{L}^4)\, .
  \label{4.13}
\end{equation}
The negative sign of this result indicates jet enhancement, i.e.,
the medium-dependence leads for small $L$ to {\it harder} jet 
fragmentation functions. Shortly after the discovery of the EMC effect, 
Nachtmann and Pirner~\cite{NP87} proposed a phenomenological 
picture which predicts harder fragmentation functions. To our
knowledge, Eq. (\ref{4.13}) is the first indication for such
a jet enhancement effect in a QCD-motivated calculation. It will
be discussed quantitatively in sections~\ref{sec4b}-\ref{sec4e}.

\subsubsection{QCD dipole phenomenology}
\label{sec4a3}

The rescattering parameter $n_0\, C$ is the only model-dependent
input of the medium-dependent gluon radiation spectrum (\ref{3.11}).
It characterizes the average squared transverse momentum transfered
from the medium to the projectile per unit pathlength, see Eq. 
(\ref{4.10}). We study this spectrum for four different values
of $n_0\, C$:
\begin{enumerate}
  \item $n_0\, C = 0.1\, {\rm fm}^{-3} = (63\, {\rm MeV})^2/ {\rm fm}$
  \item $n_0\, C = 0.5\, {\rm fm}^{-3} = (141\, {\rm MeV})^2/ {\rm fm}$
  \item $n_0\, C = 1.0\, {\rm fm}^{-3} = (200\, {\rm MeV})^2/ {\rm fm}$
  \item $n_0\, C = 2.0\, {\rm fm}^{-3} = (280\, {\rm MeV})^2/ {\rm fm}$
\end{enumerate}
We chose these values after finding in exploratory calculations
that they scan the region between negligible total energy loss
($\Delta E/E < 3\, \%$ for $n_0\, C = 0.1\, {\rm fm}^{-3}$)
and very significant energy loss ($\Delta E/E > 20\, \%$ for 
$n_0\, C = 2.0\, {\rm fm}^{-3}$).

Baier, Dokshitzer, Mueller and Schiff~\cite{BDMS99} described the 
rescattering properties of the medium by the transport coefficient 
$\hat{q}$ which satisfies $\hat{q} = 2\, n_0\, C$. They estimate
$n_0\, C = 0.05\, {\rm GeV}^3 = (500\, {\rm MeV})^2/ {\rm fm}$ for 
hot matter at a temperature $T = 250$ MeV. For cold nuclear matter, 
their estimate $n_0\, C = 0.0025\, {\rm GeV}^3 
= (111\, {\rm MeV})^2/ {\rm fm}$ is based on the relation between 
the rescattering parameter $n_0\, C$ and the gluon distribution 
$x\, G(x)$. In comparison, the two lower values ($n_0\, C = 0.1\, 
{\rm fm}^{-3}$ and $n_0\, C = 0.5\, {\rm fm}^{-3}$) of our study
scan a reasonable range of possible values for cold nuclear matter,
and the larger values of $n_0\, C$ may be associated to excited 
nuclear matter.

In principle, the parameter $n_0\, C$ can be determined 
phenomenologically for cold nuclear matter from the 
medium-dependence of hard processes. One approach~\cite{BDMPS97} 
is to relate $n_0\, C$ to the parameter $\lambda_{\rm LQS}$ which
characterizes in the LQS-factorization approach~\cite{LQS94} the 
amount of transverse momentum transfered to the hard parton. A
first determination from the dijet imbalance resulted in
$\lambda_{\rm LQS}^2 = 0.05-0.1\, {\rm GeV}^2$ corresponding
to an extremely large value 
$n_0\, C = 0.3 - 0.6\, {\rm GeV}^2/{\rm fm}$~\cite{LQS94}.
More recent studies~\cite{G98,FSSM99} use the much smaller value 
$ \lambda_{\rm LQS}^2 = 0.01\, {\rm GeV}^2$. This is still 
larger than the above estimate based on the gluon distribution. 
Significant experimental and theoretical uncertainties remain.
Especially, the connection between  $n_0\, C$ and
$\lambda_{\rm LQS}$ assumes a direct connection between multiple
soft and single secondary (hard) rescattering~\cite{BDMPS97}.
As a consistency check, one should extract $n_0\, C$ also from 
observables sensitive to the multiple soft rescattering, as e.g. DIS 
nuclear structure functions~\cite{W00}. There, however, existing 
parametrizations show a non-negligible dependence of the dipole 
strength $C$ on Bjorken $x_{\rm bj}$~\cite{GBW99} or 
$\sqrt{s}$~\cite{KST99} and it becomes important to compare
the extracted parameters in the corresponding kinematical regimes.
The present work does not aim at a phenomenological evaluation of 
existing $e-A$ and $p-A$ data which takes the above considerations
into account. It rather scans the parameter space in the region 
$0.1\, {\rm fm}^{-3} <  n_0\, C < 2.0\, {\rm fm}^{-3}$
which interpolates between cold and hot nuclear matter.

\subsection{$L$-dependence of radiative energy loss}
\label{sec4b}

As long as not stated otherwise, we present the ${\bf k}_\perp$-integrated 
expressions (\ref{4.6}), (\ref{4.7}) for the kinematical boundary 
$\chi = 1$. All numerical results are for a strong coupling constant 
$\alpha_s = 1/3$ and $N_c = 3$.
%
\begin{figure}[h]\epsfxsize=12.7cm 
\centerline{\epsfbox{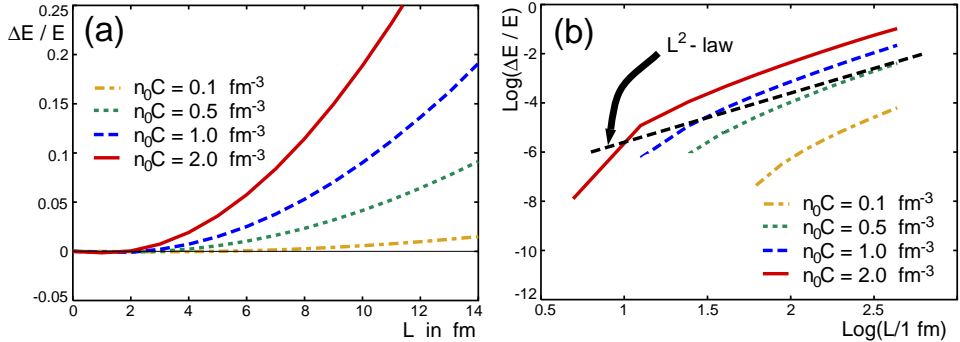}}
\vspace{0.5cm}
\caption{(a) Dependence of the medium-induced radiative energy loss 
(\protect\ref{4.7}) on the in-medium pathlength $L$ for $E = 100$ GeV,
$\chi = 1$ and different values of the rescattering parameter $n_0\, C$.
(b) Double logarithmic presentation of (a)
which indicates deviations from the BDMPS-$L^2$-law.
}\label{fig1}
\end{figure}
%
Fig.~\ref{fig1}(a) shows the
$L$-dependence of the total radiative energy loss (\ref{4.7}) 
off a hard quark of initial energy $E = 100$ GeV. 
Irrespective of the value of the rescattering parameter $n_0\, C$,
the total radiative energy loss $\Delta E/E$ increases stronger 
than linear. The double logarithmic
plot Fig.~\ref{fig1}(b) of the same calculation displays
deviations from the BDMPS-$L^2$-law~\cite{BDMPS97}. 

For sufficiently large $L$  ($L < 14\, {\rm fm}$), 
the increase of the radiative energy loss is approximately
$\propto L^{2.5}$. At smaller values of $L$ ($L < 6\, {\rm fm}$), 
the radiation spectrum shows a numerically very small increase and 
turns negativ. This is consistent 
with the analytical result (\ref{4.13}) and prevents us from extending
the double logarithmic plot Fig.~\ref{fig1}(b) into the small $L$ region.

The sensitivity of the
total radiative energy loss on the rescattering parameter $n_0\, C$ 
depends strongly on the size of the in-medium pathlength. For
$L >8$ fm, a measurement of $\Delta E/E$ with an accuracy of
$\approx 5\%$ would be sufficiently sensitive to distinguish 
between different values of the rescattering parameter $n_0\, C$. 
For smaller in-medium pathlengths,
however, this sensitivity is lost rapidly. To further analyze
the region of negative  $\Delta E/E$ and the delayed onset of
significant total radiative energy loss, we turn
now to the $x$-dependence of this spectrum.
%
\begin{figure}[h]\epsfxsize=12.7cm 
\centerline{\epsfbox{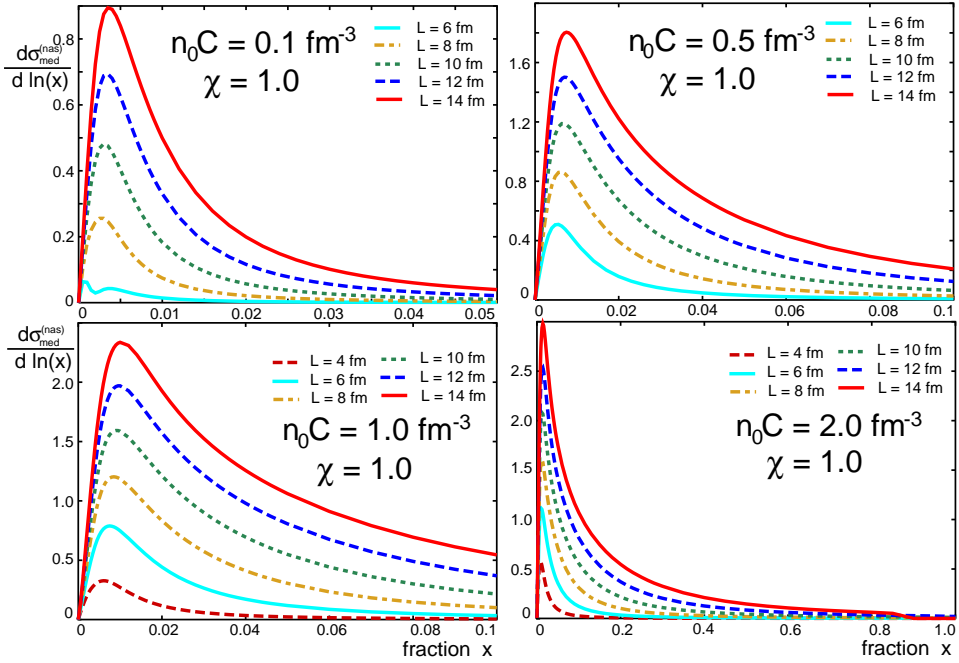}}
\vspace{0.5cm}
\caption{The $x$-differential gluon radiation spectrum 
(\protect\ref{4.6}) as a function of $x$ for different
in-medium pathlengths $L$, and different values of the
rescattering parameter $n_0\, C$.  
}\label{fig2}
\end{figure}
%
\subsection{$x$-dependence of the gluon radiation spectrum}
\label{sec4c}
Fig.~\ref{fig2} shows for relatively large in-medium
pathlengths $L$ the
$x$-differential radiation spectrum (\ref{4.6}) which enters the 
calculation of the total medium-induced radiative energy loss
(\ref{4.7}). The spectrum peaks generically
at very small values of $x$. This is an important consistency
check since the derivation of the radiation spectrum 
(\ref{2.1}) is based on the assumption $x \ll 1$. 

The maximal radiation probabilities shown in 
Fig.~\ref{fig2} exceed unity for sufficiently large pathlength $L$. 
This may be taken as an indication that multiple gluon emission plays 
an important role for sufficiently extended targets.

\begin{figure}[h]\epsfxsize=12.7cm 
\centerline{\epsfbox{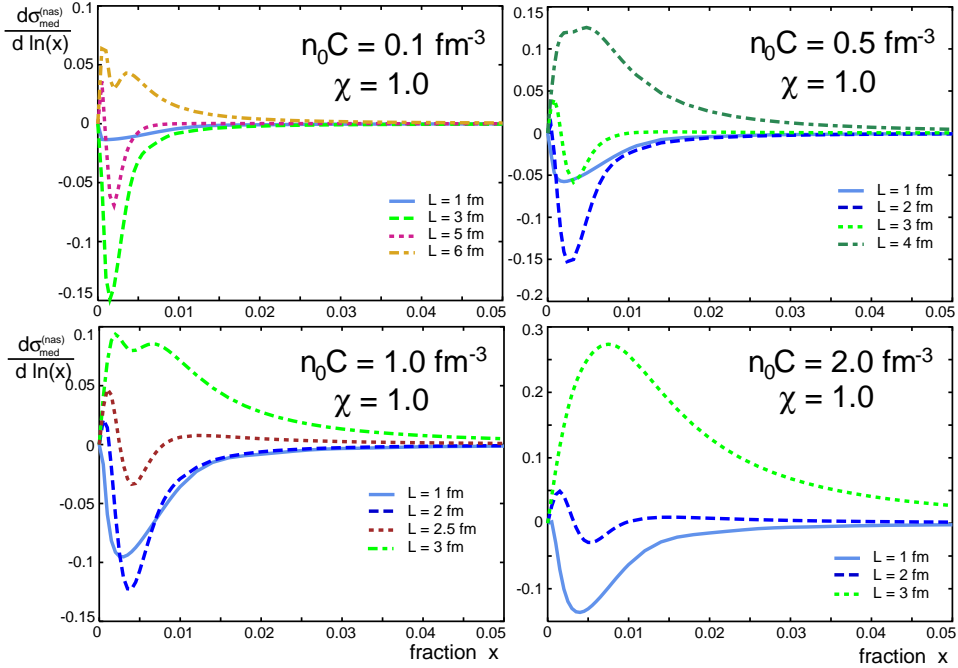}}
\vspace{0.5cm}
\caption{The same as Fig.~\protect\ref{fig2}, but for relatively
small pathlengths $L$ of the quark in the medium. The destructive
interference between hard and medium-induced radiation (``jet 
enhancement'') overcompensates the medium-induced
gluon radiation (``jet quenching'') for sufficiently small $L$.
}\label{fig3}
\end{figure}

In Fig.~\ref{fig3}, the same $x$-differential radiation spectrum 
is plotted for relatively small in-medium pathlengths.
The distinction between ``large'' and ``small'' is made
empirically by the onset of more complicated structures in the
radiation spectrum. The only purpose of presenting Fig.~\ref{fig3}
separately from Fig.~\ref{fig2} is to make these structures
better visible.
%
%
Irrespective of the value of the rescattering parameter $n_0\, C$, 
the $x$-differential medium-induced gluon radiation spectrum 
Fig.~\ref{fig3} turns negative for very small
pathlengths $L$. This is a consequence of the destructive
interference between hard and medium-induced radiation. 
As one increases $L$, more medium-induced radiation amplitudes 
become available. Whether these lead to a stronger destructive
interference with the hard radiation associated to the quark
production vertex, or whether they lead to an increasing gluon 
emission probability
depends on the quantitative interplay between in-medium pathlength
$L$ and the rescattering property $n_0\, C$ of the medium. For
$n_0\, C = 0.1\, {\rm fm}^{-3}$ in the upper right panel of Fig.~\ref{fig3},
we observe e.g. that the destructive interference effect
increases up to $L < 4$ fm. Only for larger 
pathlengths, the medium-induced gluon production probability finally 
overcompensates the destructive interference term and the
$x$-differential and $x$-integrated medium-induced energy loss
turns positive. 

According to Fig.~\ref{fig3}, the pathlength at which exact 
compensation between medium-induced radiation and destructive 
interference occurs shows a strong $n_0\, C$-dependence. It
decreases from $L \approx 5$ fm at $n_0\, C = 0.1\, {\rm fm}^{-3}$ via 
($L\approx 3$ fm, $n_0\, C = 0.5\, {\rm fm}^{-3}$) and ($L\approx 2.5$ fm, 
$n_0\, C = 1.0\, {\rm fm}^{-3}$) up to 
($L\approx 2.0$ fm, $n_0\, C = 2.0\, {\rm fm}^{-3}$). 
This decrease with increasing $n_0\, C$ is consistent with the 
picture that a significant transverse momentum transfer to the 
gluon is needed to kick it out of the kinematical region in 
which destructive interference can occur.

Quantitatively, jet enhancement ($\Delta E < 0$) at small 
pathlengths is very small. It is almost invisible in the
$x$-integrated radiation spectrum of Fig.~\ref{fig1}. However, it   
delays the onset of a positive jet quenching contribution. 
As seen from Fig.~\ref{fig3}, this 
offset depends significantly on the rescattering parameter $n_0\, C$ 
of the medium. It will be further quantified in the next subsection.
%
\begin{figure}[h]\epsfxsize=12.7cm 
\centerline{\epsfbox{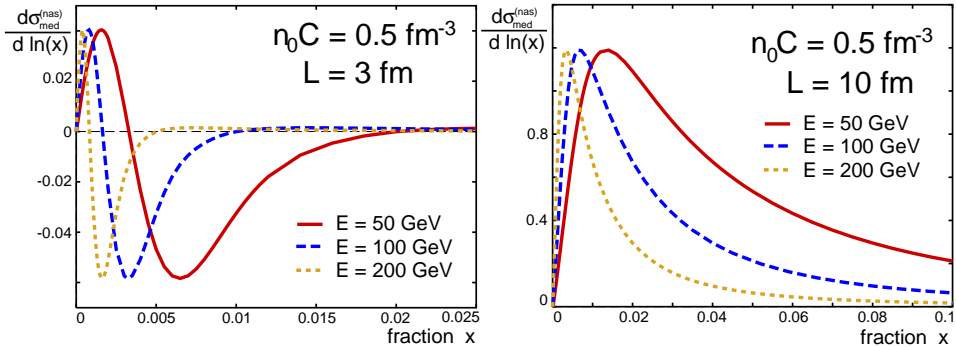}}
\vspace{0.5cm}
\caption{The energy dependence of the $x$-differential radiation
spectrum results in a simple $x$-scaling (\protect\ref{4.14})
}\label{fig4}
\end{figure}
%

\subsection{Energy dependence of the gluon radiation spectrum}
\label{sec4d}
In the gluon radiation spectrum (\ref{2.1}), the energy
$E$ of the initial quark comes always multiplied by the fraction
$x$ of the energy carried away by the gluon. This implies a trivial 
$x$-scaling of the radiation spectrum as a function of energy,
\begin{equation}
  \frac{d\sigma^{(nas)}_{med}}{d({\rm ln}x)}
        (nC,L,E,x)
        = 
        \frac{d\sigma^{(nas)}_{med}}{d({\rm ln}x)}
        (nC,L,E',\frac{E}{E'}x)\, .
 \label{4.14}
\end{equation}
In Fig.~\ref{fig4}, we show illustrative examples of this
scaling behaviour for a small pathlength (where destructive
interference effects are important) and for a large pathlength.

If the $x$-differential radiation spectrum vanishes sufficiently
fast at large $x$, the $x$-scaling (\ref{4.14}) results (as a 
function of $1/E$) in a linear increase of the area under the 
$x$-differential radiation spectrum. Then, the total radiative 
energy loss $\left[\frac{\Delta E}{E}\right]$
scales as a function of energy like
\begin{equation}
 \left[\frac{\Delta E}{E}\right] (E) = \frac{E'}{E}\, 
 \left[\frac{\Delta E}{E}\right] (E')\, .
 \label{4.15}
\end{equation}
%
\begin{figure}[h]\epsfxsize=12.7cm 
\centerline{\epsfbox{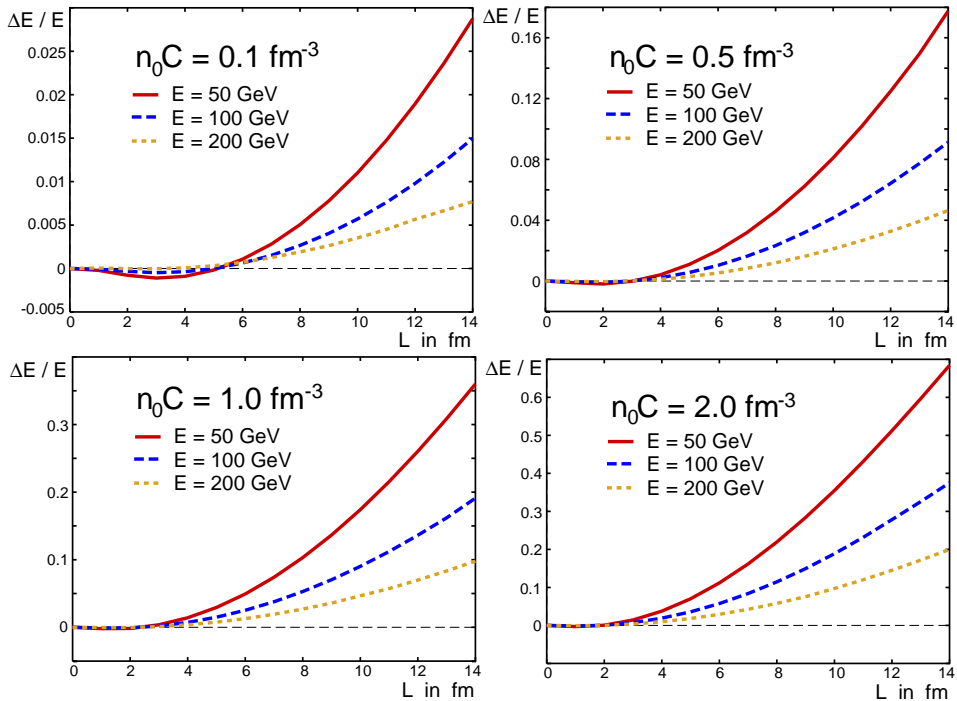}}
\vspace{0.5cm}
\caption{$L$-dependence of the fraction of medium-induced total 
radiative energy loss (\ref{4.7}) for different values of the
initial parton energy $E$ and different values of the rescattering 
parameter $n_0\, C$ of the nuclear medium.  
}\label{fig5}
\end{figure}
As can be seen from Fig.~\ref{fig5}, the value of $\Delta E/E$ doubles 
approximately if one scales the initial parton energy by a factor 
$\frac{1}{2}$. This is consistent with the scaling law (\ref{4.15}). 
It holds to good accuracy for all values of the rescattering parameter 
$n_0\, C$ and for all physically realistic pathlengths $L$: 

Fig.~\ref{fig5} also illustrates a property of the radiation spectrum
consistent with factorization:
One expects on general grounds that for increasing incident quark
energy and fixed spatial extension of the medium, the medium-induced 
changes to the parton fragmentation become negligible in the limit 
$E \to \infty$. This is the naive factorization limit in which
high momentum spectra depend on the initial parton distributions
only, and do not test properties of the medium. Fig.~\ref{fig5}
shows a significant decrease of the medium-induced fraction of
radiative energy loss $\Delta E/E$ with increasing jet energy $E$. 
However, it indicates a kinematical sweetspot where 
initial parton energies are sufficiently high ($E > 50 - 100$ GeV) 
to make jet quenching a perturbatively calculable quantity and yet 
sufficiently small ($E < 1$ TeV say) so that the dependence on
$n_0\, C$ is still significant and makes jet quenching a sensitive 
probe of non-perturbative medium properties.

Another way of looking at Fig.~\ref{fig5} is by plotting the
corresponding critical length $L_{\rm crit}$ above which the
total radiative energy loss is larger than one percent,
\begin{equation}
  \Delta E/ E (L_{\rm crit}) \equiv 0.01\, .
  \label{4.16}
\end{equation}
%
\begin{figure}[h]\epsfxsize=7.7cm 
\centerline{\epsfbox{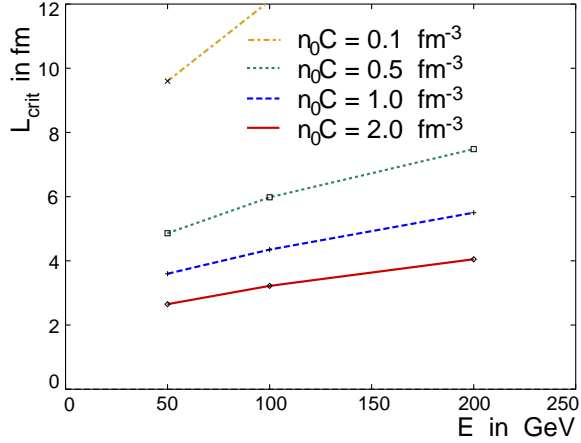}}
\vspace{0.5cm}
\caption{The critical in-medium pathlength $L_{\rm crit}$ above which
the medium-induced gluon radiation off the quark exceeds one percent
of the total quark energy 
}\label{fig6}
\end{figure}
As seen in Fig.~\ref{fig6}, $L_{\rm crit}$ increases slowly with
the incident energy of the parton. This is a consequence of the
factorization limit: for a medium of finite extension, there
is an initial parton energy above which medium-induced deviations 
of the parton fragmentation function become negligible. Also,
$L_{\rm crit}$ increases obviously if the medium transfers
less transverse momentum to the rescattering partion. $L_{\rm crit}$ 
ranges typically between $3$ fm and more than $6$ fm. 
This is a consequence of the
interplay between jet enhancement at small $L$ and jet quenching
at large $L$ which shifts the onset of a significant radiative
energy loss to larger values of the in-medium pathlength $L$.
Taking the finite geometry of ultrarelativistic heavy ion collisions
into account~\cite{LS00}, this effect may lead to a significant reduction of
the jet quenching signal at RHIC and LHC since $L_{\rm crit}$ is 
of the same order as the in-medium pathlength of a substantial fraction 
of the produced jets.
%
\subsection{Angular contribution to the radiative energy loss    
radiation spectrum}
\label{sec4e}
The $\chi$-dependence of the radiation spectrum (\ref{4.6}) allows
to calculate the gluon radiation within an angular segment 
$\Delta {\bf k}_\perp\in [|{\bf k}_{\perp 1}|,|{\bf k}_{\perp 2}|]$, 
specified by $\chi_1 = (|{\bf k}_{\perp 1}|/\omega)$ and 
$\chi_2 = (|{\bf k}_{\perp 2}|/\omega)$,
\begin{eqnarray}
  &&\frac{d\sigma^{(nas)}_{med}}{d({\rm ln}x)}
        (nC,L,E,\Delta {\bf k}_\perp)
        = \nonumber\\
        && \frac{d\sigma^{(nas)}_{med}}{d({\rm ln}x)}(nC,L,E,\chi_2)
        - \frac{d\sigma^{(nas)}_{med}}{d({\rm ln}x)}(nC,L,E,\chi_1)\, .
 \label{4.17}
\end{eqnarray}
%
\begin{figure}[h]\epsfxsize=10.7cm 
\centerline{\epsfbox{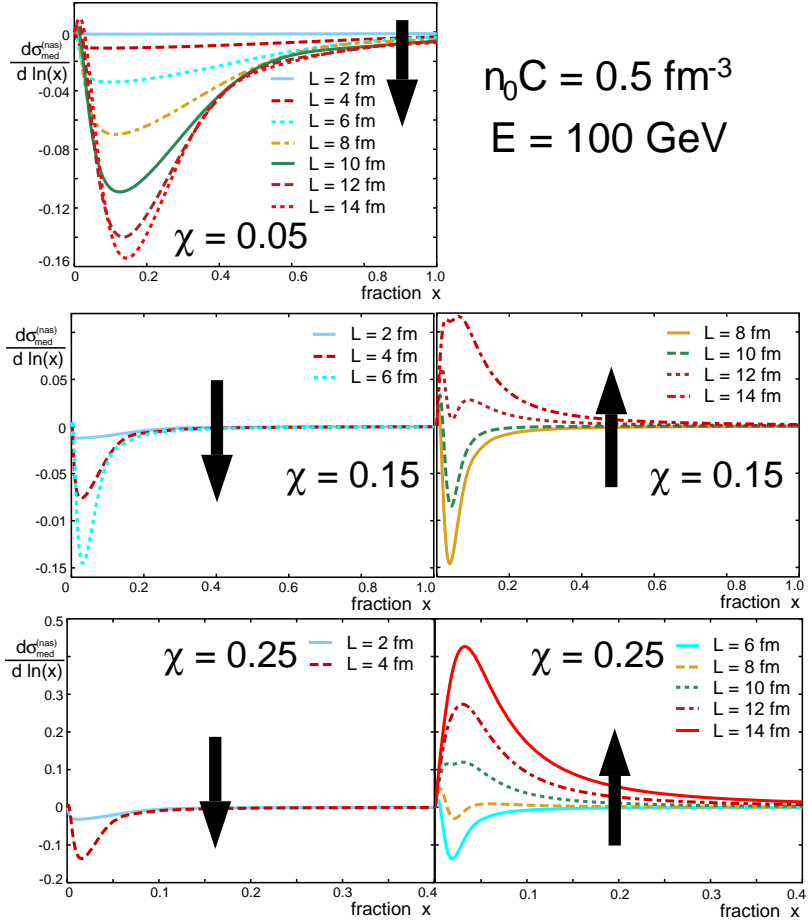}}
\vspace{0.5cm}
\caption{ The contributions to the ${\bf k}_\perp$-integrated
radiation spectrum, coming from the phase space region
$|{\bf k}_\perp| \in [0, \chi\, \omega]$ for $\chi = 0.05$,
$0.15$ and $0.25$. Arrows indicate how the spectrum changes
with increasing in-medium pathlength $L$.
}\label{fig7}
\end{figure}
%
In Fig.~\ref{fig7}, this spectrum is shown for $\chi_1 = 0$
and small values of $\chi_2 = \chi \ll 1$. For very small transverse
momenta $|{\bf k}_\perp| < 0.05\, \omega$, we find that the radiation
spectrum is significantly depleted over the whole range of available
gluon energies $x \in [0,1]$. This depletion continues to increase 
with increasing path length $L$ even for rather large L, $L = 14$ fm.
Increasing the upper bound on the transverse momentum 
$|{\bf k}_\perp| < \chi\, \omega$ to $\chi = 0.15$ and $\chi = 0.25$,
one observes that the depletion turns into an enhancement for 
sufficiently large pathlength. 

Even for large $x$ ($x > 0.5$), a significant
depletion of the spectrum occurs at small transverse momenta 
($\chi = 0.05$) but disappears if higher transverse momenta
are included ($\chi = 0.15$). This cannot be the result of
destructive interference since it does not show up in the 
${\bf k}_\perp$-integrated radiation spectrum. It comes from 
the shifting
of the gluon transverse phase space distribtion which can be
understood in terms of a classical multiple soft rescattering. 
In a classical picture, one expects a 
characteristic broadening of the hard radiation (\ref{3.11}),
\begin{equation}
        \frac{1}{{\bf k}^2_\perp} \longrightarrow
         \frac{1}{({\bf k}_\perp+ {\bf q}_\perp)^2}
        \label{4.18}
\end{equation}
as a function of the ${\bf q}_\perp$-momentum transfer which
grows by Brownian motion $\langle {\bf q}_\perp^2\rangle \propto L$.
This leads to a depletion of the low ${\bf k}_\perp$ momentum region
of the radiation spectrum without affecting the ${\bf k}_\perp$-integrated
spectrum. In contrast, the destructive interference between hard
and medium-induced radiation discussed in section~\ref{sec4c} and
also present in Fig.~\ref{fig7} affects the ${\bf k}_\perp$-integrated
spectrum.

Comparing the absolute values of the radiation spectrum for $\chi = 0.05$,
$0.1$ and $0.15$ shown in Fig.~\ref{fig7} with the absolute values of the
$\chi =1$ radiation spectrum in Figs.~\ref{fig2} and \ref{fig3}, one finds 
that the main contribution to the radiative energy loss (\ref{4.7}) comes 
from gluons emitted at relatively large transverse momenta 
$|{\bf k}_\perp| > 0.25\, \omega$. However,
the derivation of the gluon radiation spectrum (\ref{2.1}) 
assumes $|{\bf k}_\perp| \ll \omega$~\cite{BDMPS97,Z98,W00qcd}. 
We conclude that the 
calculation of the BDMPS-Z total radiative energy loss receives 
its main contribution from a kinematical region in which the validity
of the BDMPS-Z approximation is not guaranteed. This is not a 
feature of the dipole approximation studied here: numerical results 
of Gyulassy, Levai and Vitev~\cite{GLV00} indicate that also in the
opacity expansion the dominant contribution to the radiative energy 
loss is obtained from gluons with significant transverse momentum.
%
\subsection{Radiative energy loss outside a finite jet cone}
\label{sec4f}
\begin{figure}[h]\epsfxsize=12.7cm 
\centerline{\epsfbox{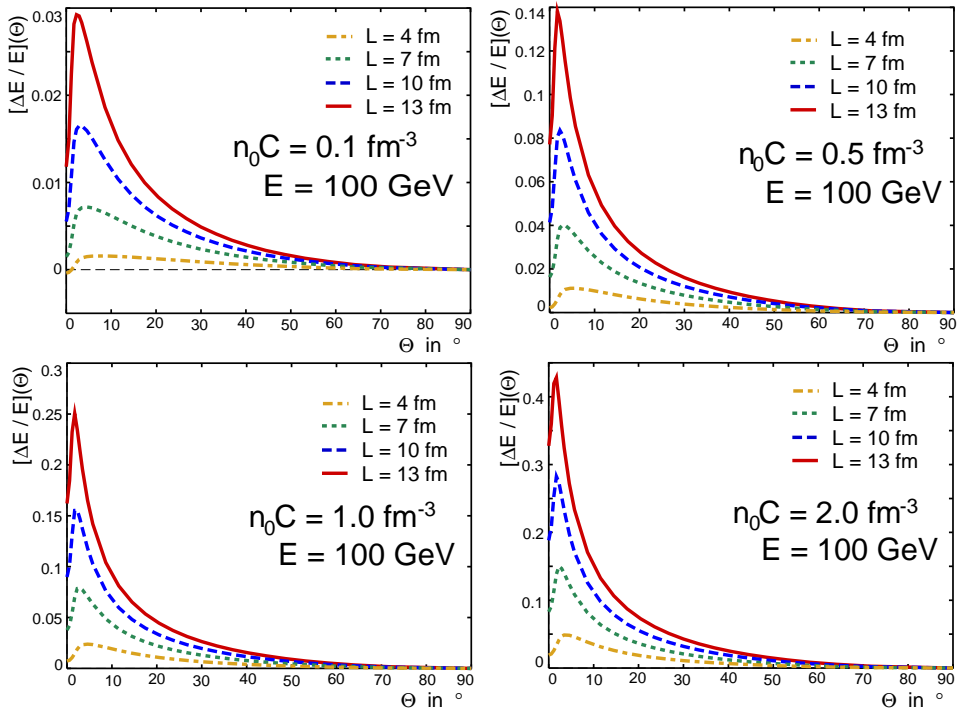}}
\vspace{0.5cm}
\caption{The fraction (\protect\ref{4.19}) of the total radiative 
energy loss $\Delta E/E$ emitted outside a jet cone of fixed angle 
$\Theta$.
}\label{fig8}
\end{figure}

\begin{figure}[h]\epsfxsize=8.7cm 
\centerline{\epsfbox{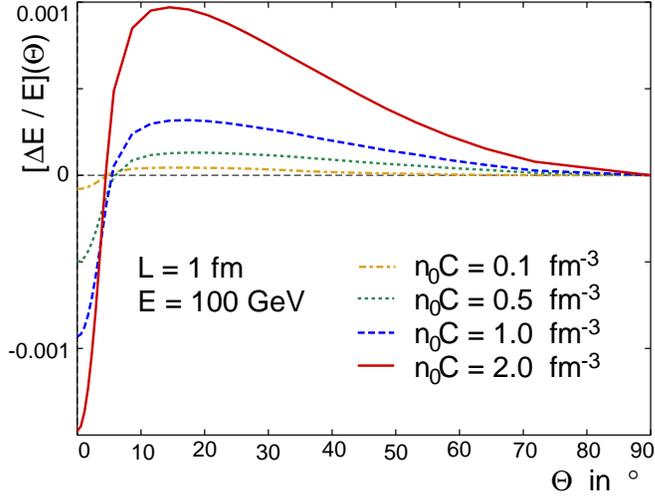}}
\vspace{0.5cm}
\caption{Same as Fig.~\protect\ref{fig8}, but for small in-medium
pathlength $L = 1$ fm.
}\label{fig9}
\end{figure}

From the $\chi$-dependence of the $x$-integrated 
radiative energy loss (\ref{4.7}), we can calculate the quantity
\begin{eqnarray}
        \frac{\Delta E}{E}(\Theta=\arcsin(\bar\chi)) = 
        \frac{\Delta E}{E}(\chi=1) - \frac{\Delta E}{E}(\bar{\chi})
        \label{4.19}\, .
 \end{eqnarray}
This describes the fraction of the medium-induced relative 
radiative energy loss $\frac{\Delta E}{E}$ radiated outside
a cone of opening angle $\Theta$. Our notation in (\ref{4.19}) 
suppresses the dependence of $ \frac{\Delta E}{E}(\Theta)$ on 
the rescattering parameter $n_0\, C$, the in-medium pathlength $L$, 
the quark energy $E$ and the fraction $x$ of the energy taken 
away by the gluon. Numerical results for (\ref{4.19}) are shown
in Fig.~\ref{fig8} for relatively large in-medium pathlengths
and in Fig.~\ref{fig9} for the small pathlengths for which the 
destructive interference between hard and medium-induced gluon
production is significant.
%
As seen from Fig.~\ref{fig8}, $\frac{\Delta E}{E}(\Theta)$ does
not decrease monotonously with increasing $\Theta$ but has a
maximum at finite jet opening angle. The reason is that the
radiative energy loss outside a cone angle $\Theta$ receives
additional contributions from the Brownian ${\bf k}_\perp$-broadening
(\ref{4.18}) of the hard vacuum radiation term $I_6$.
Such contributions do not affect the total ${\bf k}_\perp$-integrated 
yield $ \frac{\Delta E}{E}(\Theta=0)$, since they result only in 
a shifting of the transverse momentum phase space distribution of
the emitted gluon. However, this shift in transverse phase space
shows up as soon as a finite cone size is chosen. We conclude from 
Fig.~\ref{fig8} that the total ${\bf k}_\perp$-integrated radiative 
energy loss$ \frac{\Delta E}{E}(\Theta=0)$
is not the upper bound for the radiative energy loss outside a
finite jet cone angle $ \frac{\Delta E}{E}(\Theta)$. 
Depending on the rescattering parameter
$n_0\, C$ and the in-medium pathlength, the latter can be
larger by more than a factor $2$.

Fig.~\ref{fig9} displays the angular dependence (\ref{4.19})
for a small in-medium pathlength where destructive interference
becomes important. For the total ${\bf k}_\perp$-integrated radiative 
energy loss$ \frac{\Delta E}{E}(\Theta=0)$, the figure shows the small 
negative values discussed above and consistent with the analytical
result (\ref{4.13}). Outside a small opening angle 
$\Theta \approx 5^\circ$, however, the medium-induced radiative
energy loss turns positive. This is again the effect of Brownian
${\bf k}_\perp$-broadening which populates the high transverse momentum
region. 
%
\section{Conclusion}
\label{sec5}
In the above analysis of the medium-induced one-gluon radiation 
spectrum (\ref{3.11}), we found that the hard radiation component 
associated to the in vacuum production of a hard quark is 
affected by i) medium-induced radiation and 
ii) medium-induced rescattering. Both effects are non-negligible
in a quantitative analysis.

The first phenomenon is the destructive interference between 
hard (vacuum) and medium-induced radiation. For small in-medium
pathlength, analytical and numerical results indicate that the 
hard parton radiates less energy in the medium than in the vaccum. 
This jet enhancement effect is numerically small. At larger pathlength, 
it is overcompensated by a positive medium-induced radiative energy 
loss. However, a quantitatively important remainder of jet enhancement
is the delayed onset of significant jet quenching. In comparison to an 
unmodified $L^2$-law, the hard parton has to propagate through a medium 
of larger spatial extension $L > L_{\rm crit}$ in order to loose a 
detectable fraction of its total energy. In relativistic heavy ion 
collisions, the finite geometry limits the maximal extension
of the in-medium pathlength. The values of $L_{\rm crit}$ between
3 and 6 fermi found in our analysis, may thus imply
a significant reduction of the observed jet quenching signal. 

The second phenomenon is the Brownian ${\bf k}_\perp$-broadening
of the hard radiation due to multiple scattering. This shifts the
${\bf k}_\perp$-distribution of the hard (vacuum) component to 
higher transverse momenta. Subtracting the vacuum contribution, 
one finds that the fraction of the total radiative energy loss 
$R(\Theta)$ in (\ref{4.19}) emitted outside a jet angle 
$\Theta$ can be more than a factor $2$ larger than the total
medium-induced radiative energy loss obtained for $\Theta = 0^\circ$.
In contrast to naive expectation, the total ${\bf k}_\perp$-integrated 
medium-induced radiative energy loss is not an upper bound on the
medium-induced radiation outside a finite cone.
In relativistic heavy ion collisions, this rescattering effect
may imply a significant enhancement of the jet quenching signal.

Quantitative statements about the gluon radiation spectrum depend
on the energy $E$ of the projectile parton, the in-medium pathlength $L$
and the rescattering parameter $n_0\, C$ which is the only model-parameter
in our discussion. For the spectrum (\ref{3.11}), we have studied these
dependencies in detail. The suggested analysis strategy for experimental data
is to determine $n_0\, C$ for normal cold nuclear matter phenomenologically
and then to compare to the values extracted in relativistic heavy ion 
collisions. The rescattering parameter can be related in model 
calculations to the 
non-perturbative rescattering properties of the hot or cold nuclear medium. 
In this sense, $n_0\, C$ is a model parameter which can be extracted
from experimental data and which allows to distinguish 
between different physical pictures of the excited nuclear medium.

The present work is restricted to the analysis of the one-gluon 
radiation spectrum (\ref{3.11}) derived in the BDMPS-Z formalism.
It was one purpose of our quantitative study to clarify the
kinematical region of validity of this approach. In particular,
we found that the small-$x$ gluon emission probabilities are very high.
This indicates the importance of multiple gluon
radiation which is not contained in the BDMPS-Z formalism. Moreover,
the BDMPS-Z formalism is based on the assumption of small transverse 
gluon momentum $|{\bf k}_\perp| \ll \omega$ while we find the main 
contribution to radiative energy loss for $|{\bf k}_\perp| = O(\omega)$.
Both features question the validity of the BDMPS-Z formalism and 
await further work. We emphasize, however, that the 
physical origin of the two qualitatively novel effects discussed here, 
appears to be very generic. Even outside the BDMPS-Z formalism, we expect 
that the destructive interference between hard and medium-induced
radiation, and the strong implications of Brownian ${\bf k}_\perp$-
broadening for radiation outside a finite jet cone, leave similar,
quantitatively important traces in the medium-induced gluon
radiation spectrum.

\appendix
\section{Simplifying the radiation spectrum (\ref{3.11})}
\label{appa}

This appendix contains details of how to turn the radiation
cross section (\ref{2.1}) into a numerically accessible form.
We consider a nuclear medium of homogeneous density affecting
the propagation of the hard quark in the longitudinal interval 
$[0,L]$. Replacing the in-medium propagator 
${\cal K}({\bf y},y_l; {\bf u},\bar{y}_l|\omega)$
outside $[0,L]$ by the free propagator ${\cal K}_0$, 
\begin{equation}
  {\cal K}_0({\bf y},y_l; {\bf u},\bar{y}_l|\omega)
  = \frac{\omega}{2\pi i (\bar{y}_l-y_l)}\,
  \exp\left({\frac{i\omega ({\bf y}-{\bf u})^2}{2(\bar{y}_l-y_l)}}\right)\, ,
  \label{a.1}
\end{equation}
the $\epsilon$-regularization in (\ref{2.1}) can be removed
analytically~\cite{WG99}. Replacing the
constraint ${\bf y}=0$ in (\ref{2.1}) by a representation
of the $\delta$-function, $\int d{\bf y} \frac{d{\bf p}_\perp}{(2\pi)^2}
\exp\left(-i{\bf p}_\perp\cdot{\bf y}\right)$, one finds
\begin{eqnarray}
  I_1 &=& {\rm Re} \int \frac{d^2{\bf p}_\perp}{(2\pi)^2}
          \int d{\bf u}\, e^{-i{\bf u}\cdot ({\bf k}_\perp + {\bf p}_\perp)}\,
          e^{-\frac{1}{2} n_0\, \sigma({\bf u})}
          \frac{1}{{\bf p}_\perp^2}\, 
          \label{a.2}\\
  I_2 &=& \frac{\rm Re}{\omega} \int_0^L d\bar{y}_L 
          \int \frac{d^2{\bf p}_\perp}{(2\pi)^2} \int d{\bf u}\,
          d{\bf r}\, e^{-i{\bf p}_\perp\cdot {\bf r} 
            -i{\bf k}_\perp\cdot {\bf u}}
          \nonumber \\
          && \times e^{-\frac{1}{2}(L-\bar{y}_L)\, n_0\, \sigma({\bf u})}
          \frac{{\bf p}_\perp}{{\bf p}_\perp^2}\cdot
          \frac{\partial}{\partial {\bf u}}\, 
          {\cal K}({\bf r},0; {\bf u},\bar{y}_l|\omega)\, ,
          \label{a.3}\\
   I_3 &=& 2{\rm Re} \int  \frac{d^2{\bf p}_\perp}{(2\pi)^2}
          \frac{{\bf p}_\perp\cdot {\bf k}_\perp}{{\bf p}_\perp^2\, 
                {\bf k}_\perp^2} \int d{\bf r}_1\, d{\bf r}_2\,
          e^{-i{\bf p}_\perp\cdot {\bf r}_1 
            -i{\bf k}_\perp\cdot {\bf r}_2}
          {\cal K}({\bf r}_1,0; {\bf r}_2,L|\omega)\, ,
          \label{a.4}\\
  I_4 &=&   \frac{2{\rm Re}}{4\omega^2} \int_{0}^{L} dy_L  
          \int_{y_L}^{L} d\bar{y}_L\, \int d{\bf u}\, 
          e^{ -\frac{1}{2} (L-\bar{y}_L)\, n_0\, \sigma({\bf u}) }
          \nonumber \\
      &&  \times e^{-i{\bf k}_\perp\cdot{\bf u}}  \,
          {\partial \over \partial {\bf y}}\cdot 
          {\partial \over \partial {\bf u}}\, 
          {\cal K}({\bf y}=0,y_l; {\bf u},\bar{y}_l|\omega) \, ,
          \label{a.5}\\
  I_5 &=& \frac{\rm Re}{2\omega} \int_0^L dy_L\, 
          \int d{\bf u}\, e^{-i{\bf k}_\perp\cdot {\bf u}}
          \frac{{\bf k}_\perp}{{\bf k}_\perp^2} \cdot 
          \frac{\partial}{\partial {\bf y}}\, 
          {\cal K}({\bf y}=0,y_L; {\bf u},L|\omega)\, ,
          \label{a.6}\\
  I_6 &=& \frac{1}{{\bf k}_\perp^2}\, .
          \label{a.7}
\end{eqnarray}
In the dipole approximation, these expressions take the form
\begin{eqnarray}
  I_1 &=& \int \frac{d^2{\bf p}_\perp}{(2\pi)^2} 
          \frac{1}{{\bf p}_\perp^2}\, 
          \left(2\pi\over n_0\, C\, L\right)\, 
          \exp\left[
            {-\frac{({\bf k}_\perp + {\bf p}_\perp)^2}{2\, n_0\, C\, L}}
            \right]
          \label{a.8} \\
  I_2 &=& \frac{\rm Re}{\omega} \int_0^L dy_L
          \left( \frac{A(1-B^2)}{DB + iA(1-B^2)}
          \exp\left[{- \frac{B{\bf k}_\perp^2}{4[{DB + iA(1-B^2)}]}}\right]
          \right.
          \nonumber \\
          && \qquad \times \left. 
          \left( -1 + 
              \exp\left[{\frac{{\bf k}_\perp^2}{4[{DB + iA(1-B^2)}]}
                  \frac{A}{AB+iD}}\right] \right)
              \right.
          \nonumber \\
          && + \left. 
            \frac{A}{DB + iA(1-B^2)} \frac{-iD}{AB+iD}
            \exp\left[{-\frac{{\bf k}_\perp^2}{4(D-iAB)}}\right] \right)
          \label{a.9} \\
  I_3 &=& {\rm Re} \frac{2\pi}{iA(1-B^2)} \int 
           \frac{d{\bf p}_\perp}{(2\pi)^2}
           \frac{{\bf p}_\perp \cdot {\bf k}_\perp}
             {{\bf p}_\perp^2 \, {\bf k}_\perp^2}
             \nonumber \\
        && \times
             \exp\left[\frac{i\left( B{\bf p}_\perp^2
                 + 2 {\bf p}_\perp \cdot {\bf k}_\perp
                 + B{\bf k}_\perp^2\right)}{4A(1-B^2)}\right]\, ,
             \label{a.10} \\
  I_4 &=& \frac{1}{4\omega^2}\,
  2{\rm Re} \int_0^L dy_l \int_{y_l}^L d\bar{y}_l
  \left( \frac{-4A^2D}{(D-iAB)^2} 
         + \frac{iA^3B\, {\bf k}_\perp^2}{(D-iAB)^3} \right)
       \nonumber \\
       && \times 
       \exp\left[{-\frac{{\bf k}_\perp^2}{4\, (D- i\, A\, B)}}\right]\, ,
  \label{a.11}\\
  I_5 &=& \frac{1}{\omega}\, {\rm Re} \int_0^L dz\, \frac{-i}{B_z^2}\, 
        \exp\left[{-i\frac{{\bf k}_\perp^2}{4\, A_z\, B_z}}\right]\, ,
          \label{a.12}\\
  I_6 &=& \frac{1}{{\bf k}_\perp^2}\, .
          \label{a.13}
\end{eqnarray}
The variables $A$, $B$ and $D$ introduced here have different
arguments for the different terms:
\begin{eqnarray}
  A_2 &=& A_5 = \frac{\omega \Omega}{2\sin(\Omega y_L)}\, ,\qquad  
  B_2 = B_5 = \cos(\Omega y_L)\, ,
  \label{a.14} \\
  A_3  &=& \frac{\omega \Omega}{2\sin(\Omega L)}\, ,\qquad \qquad  \quad  
  B_3 = \cos(\Omega L)\, ,
  \label{a.15} \\
  A_4  &=& \frac{\omega \Omega}{2\sin(\Omega (\bar{y}_L-y_L))}\, ,\qquad 
  B_4 = \cos(\Omega (\bar{y}_L-y_L))\, ,
  \label{a.16} \\
  D_2 &=& \frac{1}{2} n_0 C (L-y_L)\, , \qquad \quad
  D_4 = \frac{1}{2} n_0 C (L-\bar{y}_L)\, .
  \label{a.17}
\end{eqnarray}


\end{document}